\documentclass[prb,twocolumn,showpacs,floatfix,superscriptaddress]{revtex4-1}

\usepackage[utf8]{inputenc}
\usepackage[T1]{fontenc}
\usepackage[margin=2cm]{geometry}
\usepackage{float}
\usepackage{color}
\usepackage{graphicx}
\usepackage{amsmath}
\usepackage{amssymb}
\usepackage{bbm}
\usepackage{indentfirst}
\usepackage{dcolumn}
\usepackage{soul}
\usepackage{mathtools}

\usepackage{subfigure}
\usepackage{multirow}
\usepackage{setspace}

\usepackage{comment}

\usepackage{hyperref}

\begin{document}

\title{Real-space nonlocal Gilbert damping from exchange torque correlation applied to bulk ferromagnets and their surfaces}

\author{Bal\'{a}zs Nagyfalusi}
\email{nagyfalusi.balazs@ttk.bme.hu}
\affiliation{Institute for Solid State Physics and Optics, HUN-REN Wigner Research Center for Physics, Konkoly-Thege M. \'{u}t 29-33, H-1121 Budapest, Hungary}
\affiliation{Department of Theoretical Physics, Institute of Physics, Budapest University of Technology and Economics, Budafoki \'{u}t 8, H-1111 Budapest, Hungary}
\author{L\'{a}szl\'{o} Szunyogh}
\email{szunyogh.laszlo@ttk.bme.hu}
\affiliation{Department of Theoretical Physics, Institute of Physics, Budapest University of Technology and Economics, Budafoki \'{u}t 8, H-1111 Budapest, Hungary}
\affiliation{HUN-REN-BME Condensed Matter Research Group, Budapest University of Technology and Economics, Budafoki \'{u}t 8, H-1111 Budapest, Hungary}
\author{Kriszti\'an Palot\'as}
\email{palotas.krisztian@wigner.hun-ren.hu}
\affiliation{Institute for Solid State Physics and Optics, HUN-REN Wigner Research Center for Physics, Konkoly-Thege M. \'{u}t 29-33, H-1121 Budapest, Hungary}
\affiliation{Department of Theoretical Physics, Institute of Physics, Budapest University of Technology and Economics, Budafoki \'{u}t 8, H-1111 Budapest, Hungary}

\date{\today}
%%%%%%%%%%%%%%%%%%%%%%%%%%%%%%%%%
\begin{abstract}
In this work we present an \textit{ab initio} scheme based on linear response theory of exchange torque correlation, implemented into the real-space Korringa-Kohn-Rostoker (RS-KKR) framework to calculate diagonal elements of the atomic-site-dependent intrinsic Gilbert damping tensor. The method is first applied to bcc iron and fcc cobalt bulk systems. Beside reproducing earlier results from the literature for those bulk magnets, the effect of the lattice compression is also studied for Fe bulk, and significant changes for the Gilbert damping are found. Furthermore, (001)-oriented surfaces of Fe and Co are also investigated. It is found that the on-site Gilbert damping increases in the surface atomic layer and decreases in the subsurface layer, and approaches the bulk value moving further inside the magnets. Realistic atomic relaxation of the surface layers enhances the identified effects. The first-neighbor damping parameters are extremely sensitive to the surface relaxation. Despite their inhomogeneity caused by the surface, the transverse Gilbert damping tensor components remain largely insensitive to the magnetization direction.

\end{abstract}
%%%%%%%%%%%%%%%%%%%%%%%%%%
\maketitle

\section{Introduction}
It is highly demanded to understand and control the dynamical processes governing the manipulation of various magnetic textures, such as atomic chains\cite{Ujfalussy2004,Etz2015}, magnetic skyrmions\cite{Iwasaki2013,Fert2013} or domain walls\cite{Schieback2007}, which can be potentially used in future magnetic recording and logic devices. 
These processes are often described by the phenomenological Landau-Lifshitz-Gilbert (LLG) equation \cite{Landau1935,Gilbert2004},
\begin{align}
\label{eq:Gilbert}
 \frac{\partial \vec{m}_i}{\partial t} = 
-\gamma \, \vec{m}_i \times \vec{B}_i^\mathrm{eff} +  \frac{\alpha}{m_i} \vec{m}_i \times \frac{\partial \vec{m}_i}{\partial t} \,, 
\end{align}
where $\vec{m}_i$  is the magnetic moment at site $i$, $m_i=|\vec{m}_i|$ is its length, and $\gamma$ is the gyromagnetic ratio. The first term on the {\em rhs} of Eq.~\eqref{eq:Gilbert} describes the precession of $\vec{m}_i$ around the effective magnetic field $\vec{B}_i^\mathrm{eff}$, while the second term is the Gilbert damping due to the energy dissipation to the lattice. Clearly, this latter term causes the relaxation of the magnetization to its equilibrium value, which is controlled by the damping constant $\alpha$ and plays a crucial role in the realization of high-speed spintronic devices.

The Gilbert damping constant $\alpha$ can be determined experimentally from the ferromagnetic resonance (FMR) spectroscopy where the damping parameter is related to the line-width in the measured spectra\cite{Kittel1948}. FMR spectroscopy is a well-established method for bulk materials\cite{Bhagat1974,Cochran1991}, but especially in the low temperature measurement it is controversial  because the intrinsic Gilbert damping  needs to be separated from various extrinsic sources of the line-width, e.g., two-magnon scattering, eddy-current damping, radiative damping, spin-pumping, or the slow relaxer mechanism \cite{Mankovsky2013,Costa2015,Lock_1966,Dillon1959,Gloanec2009,Schoen2016}. The comparison of experimental measurement to theoretical calculations is also made difficult by the sample properties like the exact atomic structure.

From a theoretical perspective the ultimate goal is to develop a method to calculate the Gilbert damping parameters from the electronic structure of the material. In the last decades there have been several efforts to understand the damping process.
The first successful method was developed by Kambersk\'y who related the damping process to the spin-orbit coupling (SOC)  in terms of the breathing Fermi surface model \cite{Kambersky1970}, while he also proposed the spin-orbit torque correlation model \cite{Kambersky1976,Gilmore2007}. Later on several other methods were introduced such as the  spin-pumping \cite{Starikov2010} and linear-response approaches \cite{brataas2008,Ebert2011,Mankovsky2013}. A recent summary of these methods was published by Guimar\~aes \textit{et al.}\ \cite{Guimaraes2019}

Due to the increased interest in noncollinear magnetism Fähnle \textit{et al.}\cite{fahnle2006} suggested an inhomogeneous tensorial damping. The replacement of a scalar $\alpha$ by a damping matrix $\underline{\underline{\alpha}}$ means that the damping field in Eq.~(\ref{eq:Gilbert}) is no longer proportional to  the time derivative of $\vec{m}_i$, it becomes a linear function of $\partial \vec{m}_i / \partial t$.
Moreover, nonlocality of the damping process implies that the damping field at site $i$ experiences $\partial \vec{m}_j / \partial t$ for any site $j$.
The LLG equation \eqref{eq:Gilbert} is then replaced by the set of equations \cite{Thonig2018},
\begin{align}
\label{eq:Gilbert_nonloc}
 \frac{\partial \vec{m}_i}{\partial t} = 
   \vec{m}_i \times \left(-\gamma \vec{B}_i^\mathrm{eff} + \sum\limits_j \underline{\underline{\alpha}}_{ij} \frac{1}{m_j} \frac{\partial \vec{m}_j}{\partial t}\right)\,,
\end{align}
where the damping term is unfolded to pairwise contributions of strength $\underline{\underline{\alpha}}_{ij}$.
The appearance of nonlocal damping terms was evidenced for magnetic domain walls\cite{Yuan2014, Nembach2013} by linking the Gilbert damping to the gradients of the magnetization.
In NiFe, Co, and CoFeB thin films Li et al. \cite{Li2016} measured wave-number-dependent dissipation using perpendicular spin wave resonance, validating thus the idea of nonlocal damping terms.
Different analytical expressions for $\underline{\underline{\alpha}}_{ij}$ are already proposed \cite{Ebert2011,Bhattacharjee2012,Gilmore2009,Thonig2018}, and the nonlocal damping is found for bulk materials\cite{Thonig2018,Miranda2021} as well as its effect on magnon properties of ferromagnets have been discussed \cite{Lu2023}.
Recent studies went further and, analogously to the higher order spin-spin interactions in spin models, introduced multi-body contributions to the Gilbert damping \cite{Brinker2022}. 

The calculation of the Gilbert damping properties of materials has so far been mostly focused on 3D bulk magnets, either in chemically homogeneous \cite{Gilmore2007,Gilmore2010,Mankovsky2013,Thonig2014,Ebert2015,Thonig2018,Guimaraes2019} or heterogeneous (e.g.\ alloyed) \cite{Ebert2011,Mankovsky2013,Miranda2021} forms. There are a few studies available reporting on the calculation of the Gilbert damping in 2D magnetic thin films \cite{Barati2014,Costa2015,Guimaraes2019,Chen2023}, or at surfaces and interfaces of 3D magnets \cite{Thonig2014,Barati2014,Miranda2021}. The calculation of the Gilbert damping in 1D or 0D magnets is, due to our knowledge, not reported in the literature.
Following the trend of approaching the atomic scale for functional magnetic elements in future spintronic devices, the microscopic understanding of energy dissipation through spin dynamics in magnets of reduced dimensions is inevitable and proper theoretical methods have to be developed.
 
Our present work proposes a calculation tool for the diagonal elements of the non-local intrinsic Gilbert damping tensor covering the 3D to 0D range of magnetic materials on an equal footing, employing a real-space embedding Green's function technique \cite{Lazarovits2002}. For this purpose, the linear response theory of the Gilbert damping obtained by the exchange torque correlation is implemented in the real-space KKR method. As a demonstration of the new method, elemental Fe and Co magnets in their 3D bulk form and their (001)-oriented surfaces are studied in the present work. Going beyond comparisons with the available literature, new aspects of the Gilbert damping in these materials are also reported.

The paper is organized as follows. In Sec.\ \ref{sec:method} the calculation of the Gilbert damping parameters within the linear response theory of exchange torque correlation using the real-space KKR formalism is given. Sec.\ \ref{sec:results} reports our results on bulk bcc Fe and fcc Co materials and their (001)-oriented surfaces. We draw our conclusions in Sec.~\ref{sec:conc}.

\section{Method}\label{sec:method}

\subsection{Linear response theory within real-space KKR}

The multiple-scattering of electrons in a finite cluster consisting of $N_C$ atoms embedded into a 3D or 2D translation-invariant host medium is fully accounted for by the equation \cite{Lazarovits2002}
\begin{align}
\label{eq_embedding}
\boldsymbol{\tau}_{\mathrm{C}}=\boldsymbol{\tau}_{\mathrm{H}}\left[\mathbf{I}-(\mathbf{t}_{\mathrm{H}}^{-1}-\mathbf{t}_{\mathrm{C}}^{-1})\boldsymbol{\tau}_{\mathrm{H}}\right]^{-1},
\end{align}
where $\boldsymbol{\tau}_{\mathrm{C}}$ and $\boldsymbol{\tau}_{\mathrm{H}}$ are the scattering path operator matrices of the embedded atomic cluster and the host, respectively, $\mathbf{t}_{\mathrm{C}}$ and $\mathbf{t}_{\mathrm{H}}$ are the corresponding single-site scattering matrices, all in a combined atomic site ($j,k\in\{1,...,N_C\}$) and angular momentum ($\Lambda,\Lambda'\in\{1,...,2(\ell_\mathrm{max}+1)^2\}$) representation: $\boldsymbol{\tau}=\{\underline{\tau}^{jk} \}=\{\tau^{jk}_{\Lambda\Lambda'}\}$ and $\mathbf{t}=\{t^{j}_{\Lambda\Lambda'}\delta_{jk}\}$, where $\ell_\mathrm{max}$ is the angular momentum cutoff in describing the scattering events, and for simplicity we dropped the energy-dependence of the above matrices.

For calculating the diagonal Cartesian elements of the nonlocal Gilbert damping tensor connecting atomic sites $j$ and $k$ within the finite magnetic atomic cluster, we use the formula derived by Ebert \textit{et al.} \cite{Ebert2011},
\begin{align}
\label{eq_alpha}
\alpha_{jk}^{\mu\mu}= \frac{2}{\pi m^{j}_\mathrm{s}}\mathrm{Tr}\left(\underline{T}^{j}_{\mu}\underline{\tilde{\tau}}_C^{jk}\underline{T}^{k}_{\mu}\underline{\tilde{\tau}}_C^{kj}\right),
\end{align}
where $\mu\in\{x,y,z\}$, the trace is taken in the angular-momentum space and the formula has to be evaluated at the Fermi energy ($E_F$). Here, $m^{j}_\mathrm{s}$ is the spin moment at the atomic site $j$, $\tilde{\tau}_{C,\Lambda \Lambda'}^{jk}=(\tau_{C,\Lambda \Lambda'}^{jk}-(\tau_{C,\Lambda' \Lambda}^{kj})^*)/2i$,
and $\underline{T}^{j}_{\mu}$ is the torque operator matrix which has to be calculated within the volume of atomic cell $j$, $\Omega_j$: $T^{j}_{\mu;\Lambda\Lambda'}=\int_{\Omega_j} d^3r Z^{j}_{\Lambda}(\vec{r})^\times \beta\sigma_{\mu}B_{\rm xc}(\vec{r})Z^{j}_{\Lambda'}(\vec{r})$, where the notation of the energy-dependence is omitted again for simplicity. Here, $\beta$ is a standard Dirac matrix entering the Dirac Hamiltonian, $\sigma_{\mu}$ are Pauli matrices, and $B_{\rm xc}(\vec{r})$ is the exchange-correlation field in the local spin density approximation (LSDA), while $Z^{j}_{\Lambda}(\vec{r})$ are right-hand side regular solutions of the single-site Dirac equation and the superscript $\times$ denotes complex conjugation restricted to the spinor spherical harmonics only \cite{Ebert2011}. 
We should emphasize that Eq.~\eqref{eq_alpha} applies to the diagonal ($\mu\mu$) elements of the Gilbert tensor only. To calculate the off-diagonal tensor elements one needs to use, e.g., the more demanding Kubo-Bastin formula \cite{Bastin1971,Bonbien2020}.    
Note also that in noncollinear magnets the exchange field $B_{\rm xc}(\vec{r})$ is sensitive to the spin noncollinearity \cite{Ricci2019} which influences the calculated torque operator matrix elements, however, this aspect does not concern our present study including collinear magnetic states only.

Note that the nonlocal Gilbert damping is, in general, not symmetric in the atomic site indices, $\alpha_{jk}^{\mu\mu}\neq\alpha_{kj}^{\mu\mu}$, instead
\begin{align}
\alpha_{kj}^{\mu\mu}=\frac{m^{j}_\mathrm{s}}{m^{k}_\mathrm{s}}\alpha_{jk}^{\mu\mu}
\end{align}
holds true. This is relevant in the present work for the ferromagnetic surfaces. On the other hand, in ferromagnetic bulk systems $\alpha_{jk}^{\mu\mu}=\alpha_{kj}^{\mu\mu}$ since $m^{j}_\mathrm{s}=m^{k}_\mathrm{s}=m_\mathrm{s}$ for any pair of atomic sites.

In practice, the Gilbert damping formula in Eq.\ (\ref{eq_alpha}) is not directly evaluated at the Fermi energy, but a small imaginary part ($\eta$) of the complex energy is applied, which is called broadening in the following, and its physical effect is related to the scattering rate in other damping theories \cite{Gilmore2007,Barati2014,Edwards2016,Thonig2018}. Taking into account the broadening $\eta$, the Gilbert damping reads
\begin{align}
\alpha_{jk}^{\mu\mu}(\eta)=-\frac{1}{4}\left[\tilde{\alpha}_{jk}^{\mu\mu}(E_+,E_+)+\tilde{\alpha}_{jk}^{\mu\mu}(E_-,E_-)\right.\nonumber\\\left.-\tilde{\alpha}_{jk}^{\mu\mu}(E_+,E_-)-\tilde{\alpha}_{jk}^{\mu\mu}(E_-,E_+)\right],
\label{eq_alpha_eta}
\end{align}
where $E_+=E_F+i\eta$ and $E_-=E_F-i\eta$, and the individual terms are
\begin{align}
&\tilde{\alpha}_{jk}^{\mu\mu}(E_1,E_2)=  \nonumber\\ 
& \frac{2}{\pi m^{j}_\mathrm{s}}  
\mathrm{Tr}\left(\underline{T}^{j}_{\mu}(E_1,E_2)\underline{\tau}_C^{jk}(E_2)\underline{T}^{k}_{\mu}(E_2,E_1)\underline{\tau}_C^{kj}(E_1)\right)
\end{align}
with $E_{1,2}\in\{E_+,E_-\}$, and the explicitly energy-dependent torque operator matrix elements are: $T^{j}_{\mu;\Lambda\Lambda'}(E_1,E_2)=\int_{\Omega_j} d^3r Z^{j\times}_{\Lambda}(\vec{r},E_1)\beta\sigma_{\mu}B_{\rm xc}(\vec{r})Z^{j}_{\Lambda'}(\vec{r},E_2)$.

\subsection{Effective damping and computational parameters}

Eq.\ (\ref{eq_alpha_eta}) gives the broadening-dependent spatially diagonal elements of the site-nonlocal Gilbert damping tensor: $\alpha_{jk}^{xx}(\eta)$, $\alpha_{jk}^{yy}(\eta)$, and $\alpha_{jk}^{zz}(\eta)$. Since no longitudinal variation of the spin moments is considered, the two transversal components perpendicular to the assumed uniform magnetization direction are physically meaningful. Given the bulk bcc Fe and fcc Co systems and their (001)-oriented surfaces with $C_{4v}$ symmetry under study in the present work, in the following the scalar $\alpha$ refers to the average of the $xx$ and $yy$ Gilbert damping tensor components assuming a parallel magnetization with the surface normal $z$[001]-direction: $\alpha_{jk}=(\alpha_{jk}^{xx}+\alpha_{jk}^{yy})/2=\alpha_{jk}^{xx}=\alpha_{jk}^{yy}$.
From the site-nonlocal spatial point of view in this work we present results on the on-site ("$00$"), first neighbor (denoted by "$01$") and second neighbor (denoted by "$02$") Gilbert damping parameters, and an effective, so-called total Gilbert damping ($\alpha_{\mathrm{tot}}$), which can be defined as the Fourier transform of $\alpha_{jk}$ at $\vec{q}=0$. The Fourier transform of the Gilbert damping reads
\begin{align}
\alpha_{\vec{q}}&=\sum\limits_{j=0}^\infty\alpha_{0j}\exp(-i\vec{q}(\vec{r}_0-\vec{r}_j))\nonumber\\
&\approx\sum\limits_{r_{0j}\leq r_{\mathrm{max}}}\alpha_{0j}\exp(-i\vec{q}(\vec{r}_0-\vec{r}_j)) \, ,
\end{align}
where $r_{0j}=|\vec{r}_0-\vec{r}_j|$
and the effective damping is defined as
\begin{align}
\alpha_{\mathrm{tot}}=\alpha_{\vec{q}=\vec{0}}=
\sum\limits_{j=0}^\infty\alpha_{0j}\approx\sum\limits_{r_{0j}\leq r_{\mathrm{max}}}\alpha_{0j}.
\label{eq_alpha_tot}
\end{align}
Since we have a real-space implementation of the Gilbert damping, the infinite summation for both quantities is replaced by an approximative summation for neighboring atoms upto an $r_\mathrm{max}$ cutoff distance measured from site "0". Moreover, note that for bulk systems the effective damping $\alpha_{\mathrm{tot}}$ is directly related to the $\vec{q}=0$ mode of FMR experiments.

The accuracy of the calculations depends on many numerical parameters such as the number of $\vec{k}$ points used in the Brillouin zone integration, the choice of the angular momentum cutoff $\ell_\mathrm{max}$, and the spatial cutoff $r_{\mathrm{max}}$ used for calculating $\alpha_{\vec{q}}$ and $\alpha_\mathrm{tot}$.
Previous research\cite{Thonig2018} showed that the Gilbert damping heavily depends on the broadening $\eta$, so  we extended our studies to a wider range of $\eta=1$ meV to 1 eV.
The sufficient $k$-point sampling was tested at the distance of $r_{\mathrm{max}}=7a_0$ (where $a_0$ is the corresponding 2D lattice constant) from the reference site with the broadening set to $1$ mRy, and the number of $\vec{k}$ points was increased up to the point, where the 5th digit of the damping became stable. Maximally, 320$\,$400 $\vec{k}$ points were used for the 2D layered calculation but the requested accuracy was reached with 45$\,$150 and 80$\,$600 $\vec{k}$ points for bulk bcc Fe and fcc Co systems, respectively.

The choice of $\ell_\mathrm{max}$ was tested through the whole $\eta$ range for bcc Fe, and it was based on the comparison of damping calculations with $\ell_\mathrm{max}=2$ and $\ell_\mathrm{max}=3$. The maximal deviation for the on-site Gilbert damping was found at around $\eta=5$ mRy, but it was still less than 10\%. The first and second neighbor Gilbert damping parameters changed in a more significant way (by $\approx 50\%$) in the whole $\eta$ range upon changing $\ell_\mathrm{max}$, yet the effective total damping was practically unchanged, suggesting that farther nonlocal damping contributions compensate this effect.
Since $\alpha_\mathrm{tot}$ is the measurable physical quantity we concluded that the lower angular momentum cutoff of $\ell_\mathrm{max}=2$  is sufficient to be used further on.

The above choice of $\ell_\mathrm{max}=2$ for the angular momentum cutoff, the mathematical criterion of positive-definite $\alpha_{jk}$ (which implies $\alpha_{\vec{q}}>0$ for all $\vec{q}$ vectors), and the prescribed accuracy for the effective Gilbert damping in the full considered $\eta=1$ meV to 1 eV range set $r_\mathrm{max}$ to 20$\, a_0$ for both bcc Fe and fcc Co. It is worth mentioning that the consideration of lattice symmetries made possible to decrease the number of atomic sites in the summations for calculating $\alpha_{\vec{q}}$ and $\alpha_{\mathrm{tot}}$ by an order of magnitude.

\section{Results and discussion}\label{sec:results}

Our newly implemented method was employed to study the Gilbert damping properties of Fe and Co ferromagnets in their bulk and (001)-oriented surfaces. In these cases only unperturbed host atoms form the atomic cluster, and the so-called self-embedding procedure \cite{Palotas2003} is employed, where Eq.~\eqref{eq_embedding} reduces to $\boldsymbol{\tau}_{\mathrm{C}}=\boldsymbol{\tau}_{\mathrm{H}}$ for the 3D bulk metals and 2D layered metal-vacuum interfaces.

\subsection{Bulk Fe and Co ferromagnets}

\begin{figure}[h!] 
\  \hspace{1.8cm} a) \hfill \  
\vspace{-1.1cm}

\includegraphics[width=0.95\columnwidth]{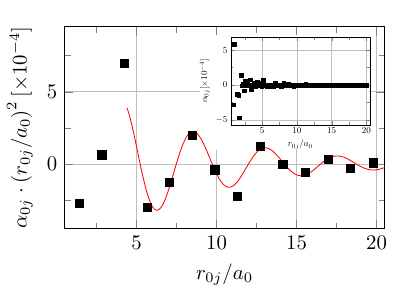}
\vspace{0.5cm}

\  \hspace{1.8cm} b) \hfill \  
\vspace{-1.1cm}

\includegraphics[width=0.95\columnwidth]{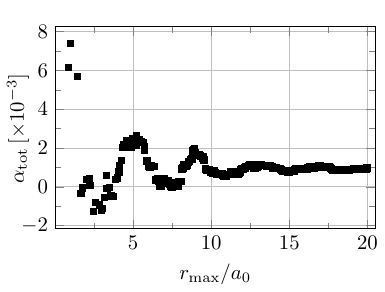}
 \caption{
  a) Nonlocal Gilbert damping in bulk bcc Fe as a function of distance $r_{0j}$ between atomic sites "0" and "$j$" shown upto a distance of 20$\, a_0$ (the 2D lattice constant is $a_0=2.863$ \AA): the black squares are calculated $\alpha_{0j}$ values times the normalized squared-distance along the [110] crystallographic direction, and the red line is the corresponding fitted curve based on Eq.\ (\ref{eq:rfuggofit}). The inset shows the nonlocal Gilbert damping $\alpha_{0j}$ values in the given distance range. b) Convergence of the effective damping parameter $\alpha_{\rm tot}$, partial sums of $\alpha_{0j}$ upto $r_{\mathrm{max}}$ based on Eq.\ (\ref{eq_alpha_tot}), where $r_{\mathrm{max}}$ is varied. The broadening is chosen to be $\eta=$68 meV.}
\label{fig:fe_rfuggo}
\end{figure}

First we calculate and analyze the nonlocal and effective dampings for bulk bcc Fe by choosing a 2D lattice constant of $a_0=2.863\,$\AA. The magnitude of the magnetic moments are obtained from the self-consistent calculation. The spin and orbital moments are  $m_\mathrm{s}=2.168\,\mu_\mathrm{B}$ and $m_\mathrm{o}=0.046\,\mu_\mathrm{B}$, respectively. The broadening is set to $\eta=68\,$meV.
The inset of Fig.\ \ref{fig:fe_rfuggo}a) shows the typical function of the nonlocal Gilbert damping $\alpha_{0j}$ depending on the normalized distance $r_{0j}/a_0$ between atomic sites "0" and "$j$". In accordance with Ref. \onlinecite{Thonig2018} the nonlocal Gilbert damping quickly decays to zero with the distance, and can be well approximated with the following function:
\begin{align}
 \alpha(r) \approx A \frac{\sin\left(kr+\phi_0\right)}{r^2}\exp(-\beta r).
 \label{eq:rfuggofit}
\end{align}
To test this assumption we assorted the atomic sites lying in the [110] crystallographic direction and fitted Eq.\ (\ref{eq:rfuggofit}) to the calculated data. In practice, the fit is made on the data set of $\alpha_{0j}(r_{0j}/a_0)^2$, and is plotted in Fig.\ \ref{fig:fe_rfuggo}a). Although there are obvious outliers in the beginning, the magnitude of the Gilbert damping asymptotically follows the $\propto\exp(-\beta r)/r^2$ distance dependence. The physical reason for this decay is the appearance of two scattering path operators (Green's functions) in the exchange torque correlation formula in Eq.\ (\ref{eq_alpha}) being broadened due to the finite imaginary part of the energy argument.

In our real-space implementation of the Gilbert damping, an important parameter for the effective damping calculation is the real-space cutoff $r_\mathrm{max}$ in Eq.\ (\ref{eq_alpha_tot}). Fig.\ \ref{fig:fe_rfuggo}b) shows the evolution of the effective (total) damping depending on the $r_{\mathrm{max}}$ distance, within which all nonlocal damping terms $\alpha_{0j}$ are summed up according to Eq.\ (\ref{eq_alpha_tot}). An oscillation can similarly be detected as for the nonlocal damping itself in Fig.\  \ref{fig:fe_rfuggo}a), and this behavior was fitted with a similar exponentially decaying oscillating function as reported in Eq.\ (\ref{eq:rfuggofit}) in order to determine the expected total Gilbert damping $\alpha_\mathrm{tot}$ value in the asymptotic $r\rightarrow\infty$ limit. In the total damping case it is found that the spatial decay of the oscillation is much slower compared to the nonlocal damping case, which makes the evaluation of $\alpha_\mathrm{tot}$ more cumbersome.
Our detailed studies evidence that for different broadening $\eta$ values the wavelength of the oscillation stays the same but the spatial decay becomes slower as the broadening is decreased (not shown). This slower decay together with the fact that the effective (total) damping value itself is also decreasing with the decreasing broadening results that below the 10\,meV range of $\eta$ the amplitude of the oscillation at the distance of 20$\, a_0$ is much larger than its asymptotic limit.
In practice, since the total damping is calculated as the $r\rightarrow\infty$ limit of such a curve as shown in Fig.\ \ref{fig:fe_rfuggo}b), this procedure brings an increased error for $\alpha_\mathrm{tot}$ below $\eta=10$ meV, and this error could only be reduced by increasing the required number of atomic sites in the real-space summation in Eq.\ (\ref{eq:rfuggofit}).

\begin{figure}[ht!]
\centering
  \includegraphics[width=1.00\columnwidth]{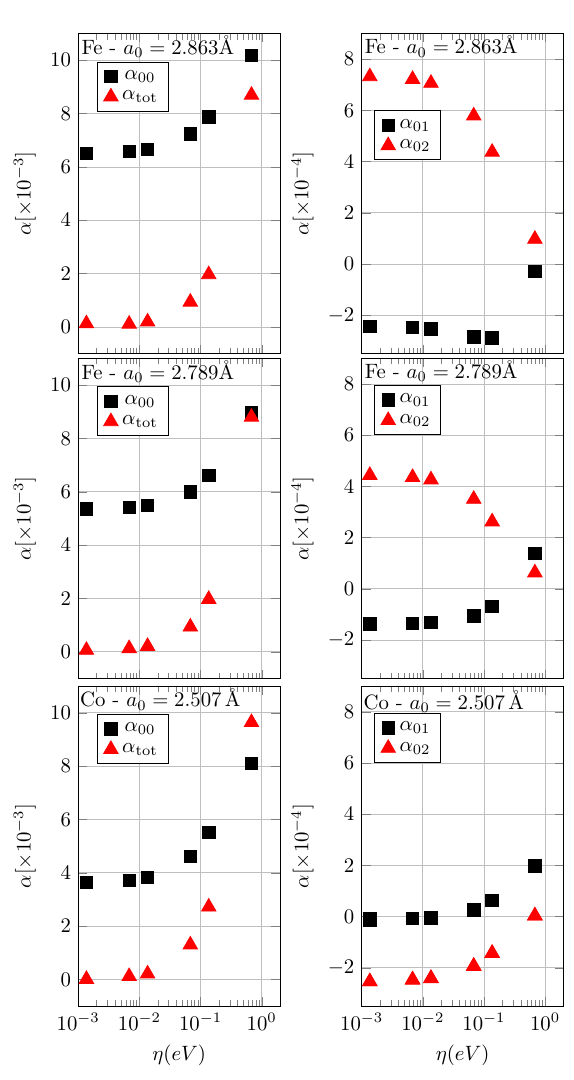}
    \caption{\label{fig:broadening_dependence}  Left column: Local on-site ($\alpha_{00}$, black square) and total ($\alpha_\mathrm{tot}$, red triangle) Gilbert damping as a function of the broadening $\eta$ for bcc Fe(001) with $a_0=2.863$\,\AA, bcc Fe(001) with $a_0=2.789$\,\AA, and fcc Co(001) with $a_0=2.507$\,\AA. Right column: Nonlocal first nearest neighbor ($\alpha_{01}$, black square) and second nearest neighbor ($\alpha_{02}$, red triangle) Gilbert damping for the same systems. 
    }
 \end{figure}

Fig.\ \ref{fig:broadening_dependence} shows the dependence of the calculated on-site, first- and second-neighbor and effective total Gilbert damping parameters on the broadening $\eta$. The left column shows on-site ($\alpha_{00}$) and total ($\alpha_\mathrm{tot}$) while the right one the first ($\alpha_{01}$) and second ($\alpha_{02}$) neighbor Gilbert dampings. We find very good agreement with the earlier reported results of Thonig et al. \cite{Thonig2018}, particularly that the on-site damping has the largest contribution to the total damping being in the same order of magnitude, while the first and second neighbors are smaller by an order of magnitude.
The obtained dependence on $\eta$ is also similar to the one published by Thonig et al. \cite{Thonig2018}: $\alpha_{00}$ and $\alpha_\mathrm{tot}$ are increasing with $\eta$, and $\alpha_{01}$ and $\alpha_{02}$ do not follow a common trend, and they are material-dependent, see, e.g., the opposite trend of $\alpha_{02}$ with respect to $\eta$ for Fe and Co. The observed negative values of some of the site-nonlocal dampings are still consistent with the positive-definiteness of the full (infinite) $\alpha_{jk}$ matrix, which has also been discussed in Ref. \onlinecite{Thonig2018}.

The robustness of the results was tested against a small change of the lattice constant simulating the effect of an external pressure for the Fe bulk. These results are presented in the second row of Fig.\ \ref{fig:broadening_dependence}, where the lattice constant of Fe is set to $a_0=2.789\,$\AA. In this case the magnetic moments decrease to $m_\mathrm{s}=2.066\,\mu_\mathrm{B}$ and  $m_\mathrm{o}=0.041\,\mu_\mathrm{B}$. It can clearly be seen that the on-site, first and second neighbor Gilbert dampings become smaller upon the assumed 2.5\% decrease of the lattice constant, but the total damping remains practically unchanged in the studied $\eta$ range. This suggests that the magnitudes of more distant non-local damping contributions are increased.

The third row of Fig.\ \ref{fig:broadening_dependence} shows the selected damping results for fcc Co with a 2D lattice constant of $a_0=2.507$\,\AA.
The spin and orbital moments are $m_\mathrm{s}=1.654\,\mu_\mathrm{B}$ and $m_\mathrm{o}=0.078\,\mu_\mathrm{B}$, respectively.
The increase of the total, the on-site, and the first-neighbor dampings  with increasing $\eta$ is similar to the Fe case, and the on-site term dominates $\alpha_{\mathrm{tot}}$.
An obvious difference is found for the second-neighbor damping, which behaves as an increasing function of $\eta$ for Co unlike it is found for Fe.

Concerning the calculated damping values, there is a large variety of theoretical methods and calculation parameters, as well as experimental setups used in the literature, which makes ambiguous to compare our results with others. Recently, Miranda \textit{et al.} \cite{Miranda2021} reported a comparison of total and on-site damping values with the available theoretical and experimental literature in their Table S1. For bcc Fe bulk they reported total damping values in the range of 1.3--4.2$\times 10^{-3}$ and for fcc Co bulk within the range of 3.2--11$\times 10^{-3}$, and our results fit very well within these ranges around $\eta\approx 100$ meV for Fe and for $\eta>100$ meV for Co. Moreover, we find that our calculated on-site damping values for bcc Fe are larger ($>5\times 10^{-3}$) than the reported values of Miranda \textit{et al.} (1.6$\times 10^{-3}$ and 3.6$\times 10^{-3}$), but for fcc Co the agreement with their reported total (3.2$\times 10^{-3}$) and on-site damping (5.3$\times 10^{-3}$) values is very good at our $\eta=136$ meV broadening value.

\begin{figure}[h!]
  \includegraphics[width=0.49\columnwidth]{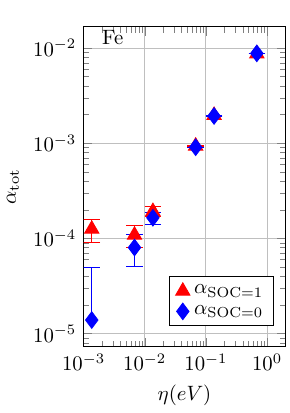}
  \includegraphics[width=0.49\columnwidth]{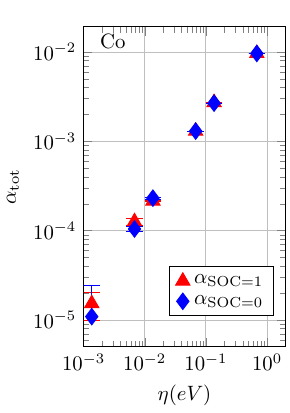}
    \caption{\label{fig:soc_fugges} Effective (total) Gilbert damping for bcc Fe (left) and  fcc Co (right) as a function of broadening $\eta$ on a log-log scale. The error bars are estimated from the fitting procedure of Eq.\ (\ref{eq:rfuggofit}). The red triangles show the case with normal SOC ($\alpha_{\mathrm{SOC}=1}$), and the blue diamonds where SOC is switched off ($\alpha_{\mathrm{SOC}=0}$).
    }
\end{figure}

Next, we investigate the spin-orbit-coupling-(SOC)-originated contribution to the Gilbert damping. Our method makes it inherently possible to include a SOC-scaling factor in the calculations\cite{Ebert1997}. Fig.\ \ref{fig:soc_fugges} shows the obtained total Gilbert damping as a function of the broadening $\eta$ with SOC switched on/off for bcc Fe and fcc Co. It can be seen that the effect of SOC is not dominant at larger $\eta$ values, but the SOC has an important contribution at small broadening values ($\eta<10^{-2}$ eV), where the calculated total Gilbert damping values begin to deviate from each other with/without SOC. As discussed in Ref. \onlinecite{Guimaraes2019}, without SOC the damping should go toward zero for zero broadening, which is supported by our results shown in Fig.\ \ref{fig:soc_fugges}.

\subsection{(001)-oriented surfaces of Fe and Co ferromagnets}

In the following, we turn to the investigation of the Gilbert damping parameters at the (001)-oriented surfaces of bcc Fe and fcc Co. Both systems are treated as a semi-infinite ferromagnet interfaced with a semi-infinite vacuum within the layered SKKR method \cite{skkr}. In the interface region 9 atomic layers of the ferromagnet and 3 atomic layers of vacuum are taken, which is sandwiched between the two semi-infinite (ferromagnet and vacuum) regions. Two types of surface atomic geometries were calculated: (i) all atomic layers having the bulk interlayer distance, and (ii) the surface and subsurface atomic layers of the ferromagnets have been relaxed in the out-of-plane direction using the Vienna Ab-initio Simulation Package (VASP) \cite{vasp} within LSDA\cite{Ceperley1980}. For the latter case the obtained relaxed atomic geometries are given in Table~\ref{tbl:surf_relax}.

\begin{table}[]
\caption{Geometry relaxation at the surfaces of the ferromagnets: change of interlayer distances relative to the bulk interlayer distance at the surfaces of bcc Fe(001) and fcc Co(001), obtained from VASP calculations. "L1" denotes the surface atomic layer, "L2" the subsurface atomic layer, and "L3" the sub-subsurface atomic layer. All other interlayer distances are unchanged in the geometry optimizations.}
   \label{tbl:surf_relax}
  \begin{tabular}{c|cc}
   \hspace{2cm}      & \hspace{0.5cm}L1-L2\hspace{0.5cm} & \hspace{0.5cm}L2-L3\hspace{0.5cm}   \\\hline
 bcc Fe(001) &  -13.7\%  & -7.7\%\\
 fcc Co(001) & -12.4\%  & -6.4\%   \\
 \end{tabular}
\end{table}

\begin{figure}[h!] \centering
\includegraphics[width=0.90\columnwidth]{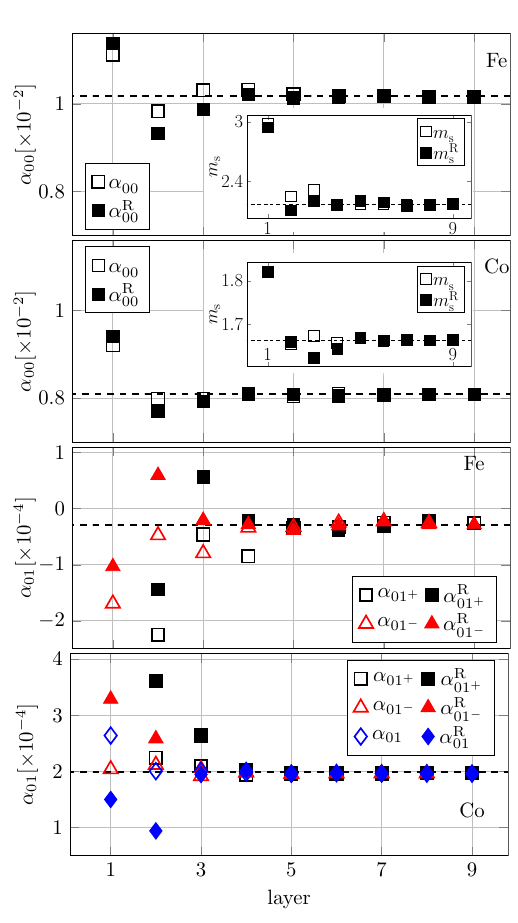}
\caption{\label{fig:layer_dep} Evolution of the layer-resolved Gilbert damping from the surface atomic layer (L1) of bcc Fe(001) and fcc Co(001) toward the bulk (L9), depending also on the out-of-plane atomic relaxation "R". On-site ($\alpha_{00}$) and first neighbor ($\alpha_{01}$) Gilbert damping values are shown in the top two and bottom two panels, respectively. The broadening is $\eta=0.68$ eV. The empty symbols belong to the calculations with the ideal bulk interlayer distances, and the full symbols to the relaxed surface geometry, denoted with index "R". Note that $\alpha_{01}$ is calculated for nearest neighbors of atomic sites in the neighboring upper, lower, and the same atomic layer (for fcc Co only), and they are respectively denoted by "$+$" (L-(L+1)), "$-$" (L-(L$-$1)), and no extra index (L-L). The insets in the top two panels show the evolution of the magnitudes of the layer-resolved spin magnetic moments $m_s$. The horizontal dashed line in all cases denotes the corresponding bulk value.}
\end{figure}

Figure \ref{fig:layer_dep} shows the calculated layer-resolved on-site and first-neighbor Gilbert damping values (with $\eta=0.68$ eV broadening) for the bcc Fe(001) and fcc Co(001) surfaces. It can generally be stated that the surface effects are significant in the first 4 atomic layers of Fe and in the first 3 atomic layers of Co. We find that the on-site damping ($\alpha_{00}$) increases above the bulk value in the surface atomic layer (layer 1: L1), and decreases below the bulk value in the subsurface atomic layer (L2) for both Fe and Co. This finding is interesting since the spin magnetic moments ($m_s$, shown in the insets of Fig.\ \ref{fig:layer_dep}) are also considerably increased compared to their bulk values in the surface atomic layer (L1), and the spin moment enters the denominator when calculating the damping in Eq.\ (\ref{eq_alpha}). $\alpha_{00}$ increases again in L3 compared to its value in L2, thus it exhibits a nonmonotonic layer-dependence in the vicinity of the surface. The damping results obtained with the ideal bulk interlayer distances and the relaxed surface geometry ("R") are also compared in Fig.\ \ref{fig:layer_dep}. It can be seen that the on-site damping is increased in the surface atomic layer (L1), and decreased in the subsurface (L2) and sub-subsurface (L3) atomic layers upon atomic relaxation ("R") for both Fe and Co. The first-neighbor dampings ($\alpha_{01}$) are of two types for the bcc Fe(001) and three types for the fcc Co(001), see caption of Fig.\ \ref{fig:layer_dep}.
All damping values are approaching their corresponding bulk value moving closer to the semi-infinite bulk (toward L9). In absolute terms, for both Fe and Co the maximal surface effect is about $10^{-3}$ for the on-site damping, and $~2\times10^{-4}$ for the first-neighbor dampings. Given the damping values, the maximal relative change is about 15\% for the on-site damping, and the first-neighbor dampings can vary by more than 100\% (and can even change sign) in the vicinity of the surface atomic layer.
Note that Thonig and Henk \cite{Thonig2014} studied layer-resolved (effective) damping at the surface of fcc Co within the breathing Fermi surface model combined with a tight-binding electronic structure approach. Although they studied a different quantity compared to us, they also reported an increased damping value in the surface atomic layer, followed by an oscillatory decay toward bulk Co.

So far the presented Gilbert damping results correspond to spin moments pointing to the crystallographic [001] ($z$)  direction, and the transverse components of the damping $\alpha^{xx}$ and $\alpha^{yy}$ are equivalent due to the $C_{4v}$ symmetry of the (001)-oriented surfaces. In order to study the effect of a different orientation of all spin moments on the transverse components of the damping, we also performed calculations with an effective field pointing along the in-plane ($x$) direction: [100] for bcc Fe and [110] for fcc Co.
In this case, due to symmetry breaking of the surface one expects an anisotropy in the damping, i.e., that the transverse components of the damping tensor, $\alpha^{yy}$ and $\alpha^{zz}$, are not equivalent any more.
According to our calculations, however, the two transverse components of the on-site ($\alpha_{00}^{yy}$ and $\alpha_{00}^{zz}$) and nearest-neighbor ($\alpha_{01}^{yy}$ and $\alpha_{01}^{zz}$) damping tensor, at the Fe surface differed by less than 0.1~\% and at the Co surface by less than 0.2~\%, i.e., despite the presence of the surface the damping tensor remained highly isotropic. The change of the damping with respect to the orientation of the spin moments in $z$ or $x$ direction (damping anisotropy) turned out to be very small as well: the relative difference in $\alpha_{00}^{yy}$ was 0.1~\% and 0.3~\%, while 0.5~\% and 0.1~\% in $\alpha_{01}^{yy}$ for the Fe and the Co surfaces, respectively. For the farther neighbors, this difference was less by at least two orders of magnitude.

\section{Conclusions}\label{sec:conc}
We implemented an \textit{ab initio} scheme of calculating diagonal elements of the atomic-site-dependent Gilbert damping tensor based on linear response theory of exchange torque correlation into the real-space Korringa-Kohn-Rostoker (KKR) framework. To validate the method, damping properties of bcc Fe and fcc Co bulk ferromagnets are reproduced in good comparison with the available literature. The lattice compression is also studied for Fe bulk, and important changes for the Gilbert damping are found, most pronounced for the site-nonlocal dampings. By investigating (001)-oriented surfaces of ferromagnetic Fe and Co, we point out substantial variations of the layer-resolved Gilbert damping in the vicinity of the surfaces depending on various investigated parameters. The effect of such inhomogeneous dampings should be included into future spin dynamics simulations aiming at an improved accuracy, e.g., for 2D surfaces and interfaces. We anticipate that site-nonlocal damping effects become increasingly important when moving toward physical systems with even more reduced dimensions (1D).

\begin{acknowledgments}
The authors acknowledge discussions with Danny Thonig. Financial support of the National Research, Development, and Innovation (NRDI) Office of Hungary under Project Nos.\ FK124100 and K131938, the János Bolyai Research Scholarship of the Hungarian Academy of Sciences (Grant No.\ BO/292/21/11), the New National Excellence Program of the Ministry for Culture and Innovation from NRDI Fund (Grant No.\ ÚNKP-23-5-BME-12), and the Hungarian State E\"otv\"os Fellowship of the Tempus Public Foundation (Grant No.\ 2016-11) are gratefully acknowledged. Further support was provided by the Ministry of Culture and Innovation of Hungary from the NRDI Fund through the grant no. TKP2021-NVA-02.
\end{acknowledgments}

\bibliography{paper_gilbert_bulk_surface}{}

\end{document}